\documentclass[final,12pt]{revtex4}
\usepackage{graphicx}
\usepackage[]{epsfig}
\begin{document}

\markboth{H. P\'{e}rez Rojas and E. Rodr\'{i}guez Querts} {MAGNETIC
FIELDS IN QUANTUM DEGENERATE SYSTEMS AND IN VACUUM}

%
%

\title{MAGNETIC FIELDS IN QUANTUM DEGENERATE SYSTEMS AND IN VACUUM}

\author{H. P\'{E}REZ ROJAS \\ E. RODR\'{I}GUEZ QUERTS}

\address{ICIMAF, Calle E No. 309, La Habana,10400, Cuba}



\begin{abstract}
We consider self-magnetization of charged and neutral vector bosons
bearing a magnetic moment in a gas and in vacuum. For charged vector
bosons (W bosons) a divergence of the  magnetization in both the
medium and the electroweak vacuum  occurs for the critical field
$B=B_{wc}=m_{w}^{2}/e$. For $B>B_{wc}$ the system is unstable. This
behavior suggests the occurrence of a phase transition at $B=B_{c}$,
where the field is self-consistently maintained. This mechanism
actually prevents $B$ from reaching the critical value $B_{c}$. For
virtual neutral vector bosons bearing an anomalous magnetic moment,
the ground state has a similar behavior for
$B=B_{nbc}=m_{nb}^{2}/q$. The magnetization in the medium is
associated to a Bose-Einstein condensate and we conjecture a similar
condensate occurs also in the case of vacuum.
 The model is applied to virtual electron-positron pairs
bosonization in a magnetic field  $B \sim B_{pc}\lesssim
2m_{e}^{2}/e$, where $m_e$ is the electron mass. This would lead
also to vacuum self-magnetization in QED, where in both cases the
symmetry breaking is due to a condensate of quasi-massless
particles.
\end{abstract}
\maketitle

\section{Introduction}
Macroscopic bodies become unstable when its rest energy is of the
same order than the interaction energy with some field. For
instance, a classical instability is produced when the
gravitational and rest energies of a body of mass $M$ and radius
$R$ are of the same order $Mc^{2}\sim GM^2/R$, leading to a
gravitational collapse. Quantum magnetic collapse for macroscopic
magnetized  objects is also claimed to occur for high magnetic
fields \cite{{aurora1},{aurora2},{Chaichian}} when the magnetic
energy density tends to be equal to the internal energy density.

  Of especial interest is the
instability of the energy ground state resulting from the solution
of the Dirac equation for an electron in a Coulomb field large
enough. The eigenvalues \cite{Schiff} are $E = m c^2/\sqrt{1 +
\frac{Z^2 \alpha^2}{(s + n')^2}},$ where $s = \pm(k^2 - Z^2
\alpha^2)^{1/2}$, and $k =\pm 1, \pm 2,...$, $n'= 0,1,2..$. It is
well known that for $n'=0$, $k=1$ the denominator diverges if
$Z=1/\alpha$. That is, there is a critical electric field
$E_c=Ze/\lambda_c^2=m^2 c^3/e \hbar$ for which electrons and
positrons can be created spontaneously from the decay of vacuum:
vacuum boils. Thus, atoms with atomic number
$Z\geq\frac{1}{\alpha}\approx137$ are not stable, due to the QED
vacuum instability for electric fields $E> E_c$.

 The usual electroweak vacuum in an external
magnetic field $B$ has also an instability for fields greater than
some critical value $B_{wc} = m_{w}^2/e\sim 10^{24}G$, due to the
presence of charged vector bosons W ($m_w$ is the W boson mass).
This problem was studied by Ambjorn and Olesen \cite{{ao1},{ao2}}
who found solutions  $B>B_{wc}$ for classical equations of motion.

Here we analyze   the problem of quantum
  stability of degenerate $W$ boson gas and vacuum in a magnetic field,
  starting from the quantum statistical point of view and methods.
  The
present authors interpret the mentioned instability of bosonic
vacuum as indicating a phase transition to a self-magnetized
state. We want to remark that for much lower fields, $B \sim
10^{20}G$ the similar instability might appear for instance, for
$\rho$, $\omega$ vector mesons or paired fermions in states of
spin unity. A similar behavior has been found in the case of
neutral vector bosons with an anomalous magnetic moment which
suggests the applicability of this model to describe the
positronium behavior in a strong magnetic field and to discuss the
possibility of QED vacuum self-magnetization.
\section{Charged vector bosons}
 For  $W$ bosons in an external magnetic
field $B_j=B\delta_{j3}$ ($j=3$) the energy spectrum  is
 \begin{equation}
 E_{wn}=\sqrt{p_{3}^{2}+m_{w}^{2}+2eB(n+\frac{1}{2})}, \hspace{1cm}
 E_{wg}=\sqrt{p_{3}^{2}+m_{w}^{2}-eB},
 \end{equation}
where $n = 0, 1, 2...$ are the Landau quantum numbers and $p_{3}$ is
the momentum component along the field direction. The ground state
$E_{wg}(p_{3}=0)$ vanishes for $B=B_{wc}$, and becomes imaginary
(unstable) for $B>B_{wc}$.

We started from the expression for the thermodynamic potential at
the tree level approximation
\begin{equation}\label{wpot}
\Omega_{w}= \Omega_{sw} + \Omega_{0w}.
\end{equation}
 The first term in (\ref{wpot}) is the
statistical contribution\cite{hugo}
\begin{eqnarray}
\Omega_{sw} = \frac{eB}{4 \pi^2 \beta} \int_{-\infty}^{\infty} dp_3
\ln [ (1 - e^{-(E_{wg} - \mu_w)\beta})(1 - e^{-(E_{wg} +
\mu_w)\beta}) ]+  \nonumber \\
 +\frac{eB}{4 \pi^2 \beta} \sum_{n = 0}^{\infty} b_n
\int_{-\infty}^{\infty} dp_3 \ln [ (1 - e^{-(E_{wn} -
\mu_w)\beta})(1 - e^{-(E_{wn} + \mu_w)\beta})].
\end{eqnarray}
Here  $b_n=3-\delta_{n0}$,  $\beta=1/T$ is the inverse of
temperature and  $\mu_w$ is the chemical potential. The second one
\begin{equation}\nonumber
\Omega_{0w} =
\frac{eB}{4\pi^2}\int_{-\infty}^{\infty}
dp_{3}(\sum_{n=0}^{\infty}E_{wg}+b_n E_{wn}),
\end{equation}
  corresponds to  the zero point energy density of vacuum, obtained
as the zero temperature and zero density limit of $\Omega_{w}$.
After regularization, we get  an Euler-Heisenberg\cite{Euler} like
term
\begin{equation}\nonumber
\Omega_{0w} = -\frac{e^{2}B^2}{16\pi^2}\int_0^{\infty}e^{-B_{wc}
x/B}(\frac{1+2cosh 2x}{sinh x} -\frac{3}{x}-\frac{7x}{2})\frac{d
x}{x^{2}}<0 \label{EHlike}.
\end{equation}

The $W$ boson density  $N_w=-\frac{\partial \Omega_{w}}{\partial
\mu_w}$ looks like
\begin{equation}
N_w=\frac{eB}{4\pi ^2}\left[ \int_{-\infty }^\infty dp_3\left(
n_{ow^{-}}-n_{ow^{+}}\right) +\sum_{n=o}^\infty \beta
_n\int_{-\infty }^\infty dp_3\left( n_{nw^{-}}-n_{nw^{+}}\right)
\right] ,
\end{equation}
\begin{equation}
n_{ow^{\pm }}=\left[ e^{\left( E _{wg}\mp \mu _w\right) \beta
}-1\right] ^{-1}, \hspace{1cm} n_{nw^{\pm }}=\left[ e^{\left(
E_{wn}\mp \mu _w\right) \beta }-1\right] ^{-1}.
\end{equation}

The magnetization ${\cal M}_{w}=-\partial \Omega_w/\partial B$
contains the contributions of both real and virtual $W$ bosons
\begin{equation}
{\cal M}_{w}={\cal M}_{sw}+{\cal M}_{0w},
 \end{equation}
 being
\begin{eqnarray}
{\cal M}_{sw}&=&-\frac{\Omega _w}B+\frac{e^2B}{4\pi ^2}\int_{-\infty }^\infty dp_3%
\frac{\left( n_{ow^{-}}+n_{ow^{+}}\right) }{\varepsilon _{ow}}-  \nonumber \\
&&-\frac{e^2B}{4\pi ^2}\sum_{n=o}^\infty \beta _n\left( n+\frac
12\right)
\int_{-\infty }^\infty dp_3\frac{\left( n_{nw^{-}}+n_{nw^{+}}\right) }{%
\varepsilon _{nw}}.  \label{wmagn}
\end{eqnarray}
It can be observed that the expression (\ref{wmagn}) contains
positive (ferromagnetic) and negative (diamagnetic) contributions,
coming from the ground and excited Landau states, respectively.
 Vacuum shows a paramagnetic behavior,
described by
\begin{equation}\label{wvacmagn}
{\cal M}_{0w}=-2\frac{\Omega_{0e}}{B}+ \frac{e
m_{w}^{2}}{16\pi^2}\int_0^{\infty}e^{-B_{wc} x/B}(\frac{1+2cosh
2x}{sinh x} -\frac{3}{x}-\frac{7x}{2})\frac{ d x}{x}>0.
\end{equation}

\subsection{Degenerate
limit}
    For $eB\gg T^2$ , the average W boson population in excited Landau states is negligible
small, and most of the W density is in the Landau ground state,
which  near the zero momentum along $\textbf{B}$ behaves as
\begin{equation}
\Omega_{sw} \sim \frac{eB}{4 \pi^2 \beta} \int_{0}^{p_0} dp_3 \ln
[({\sqrt{p_3^2 + m_w^2-eB)} - \mu_w)\beta}].
\end{equation}
If $\mu_w \to (m_w^2-eB)$, we would have an infrared divergence at
$p_3 =0$. Actually, the population in the Landau ground state
increases as the parameter $T/N_w^{1/3}$ decreases: there is a Bose
condensation but no critical temperature. For such conditions if
$N_{0w}$ is the density in the ground state, the magnetization is
\begin{equation}
{\cal M}_{sw}=eN_{0w} /2\sqrt{m_w^2-eB},
\end{equation}
and the condition of self magnetization for
$\frac{B}{B_{wc}}\equiv \eta$ can be written as
\begin{equation}\label{wstselfmagn}
    \eta=\frac{a}{ \sqrt{1-\eta}}, \hspace{1cm} a=\frac{2\pi e^2
    N_w}{m_{w}^3}.
\end{equation}
We see that for $eB \to m_w^2$, the  field can be maintained
self-consistently  for
 $N_{w} \leq \frac{m_{w}^3 a_{max}}{2 \pi e^2}\sim 10^{47}$
($a_{max}=0.3849$), the instability of the thermodynamic potential
at $\mu_w =(m_w^2-eB)=0$ due to the arising of effectively massless
vector charged particles, can be avoided and the field intensity is
kept always\cite{{ariel},{ariel2}} as $B< B_{wc} \sim m_w^2/e$. The
condensate behaves as ferromagnetic.

From the general expression foe the energy momentum
tensor\cite{Chaichian} we get anisotropic pressures $P_{w3}$ and
$P_{w\perp}$ in the directions parallel and perpendicular to the
magnetic field, respectively,
\begin{equation}\label{parpr}
P_{w3} = -\Omega_{w}
\end{equation}
\begin{equation}\label{magnpr}
P_{w\perp}=-\Omega_{w}-B{\cal M}_{w},
\end{equation}
 We see that
the contribution of observable particles, given by the statistical
term in the expression for the total thermodynamic potential,
vanishes in the zero temperature and zero chemical potential limit.
The remaining term leads to the zero point energy of vacuum.  For
vacuum,  the average particle density vanishes, but other quantities
like energy density, magnetization and pressures are non-zero.

Experimental results in condensed matter have shown \cite{Zwerlein}
that a fermion gas bosonize  for temperatures close to zero (in
usual CGS units, the adimensional parameter $T/\hbar cN^{1/3}$ small
enough), leading either to BCS pairing or to Bose-Einstein
condensation. The last case shows ferromagnetic properties
\cite{Klausen}. Thus, if the thermal disorder decreases enough, it
leads through a phase transition to a lower energy highly ordered
state. As the increasing magnetic field produces also an increasing
order of the fermion system, and if $eB\gg T^2$, the mechanism of
bosonization might lead to lower energy states through Bose-Einstein
condensation. Thus, the occurrence of such mechanism is interesting
in connection to the origin of large magnetic fields in some white
dwarfs (e.g. $10^{10}$G), if vector pairing electrons occur and
condense\cite{Hugo7}. (Also, if charged vector di-quarks are formed
in neutron stars, a version of our model might be of interest in
explaining the arising of larger fields ($>10^{13}$G) in neutron
stars).

\subsection{$W$ boson vacuum}
 In electroweak vacuum the magnetization (\ref{wvacmagn}) diverges for $B
\geq B_{wc}$. One can pick up the logarithmical infrared divergence
by considering again a neighborhood of zero momentum for the Landau
ground state. One gets the term
\begin{equation}
{\cal M}_{0w} \sim -\frac{e^2
B_{wc}}{8\pi^2}\ln{\left(\frac{B_{wc}}{B}-1 \right)} >0.
\end{equation}
This divergence (Fig. 1) is indicating a phase transition to a
ferromagnetic state for $B \approx B_{wc}$, which we may understand
as due to a sort of Bose-Einstein condensation of electrically
positive and negative virtual quanta whose effective mass is
non-zero but arbitrary small.

By equating $B=4\pi {\cal M}_{vac}$, one can obtain an electroweak
vacuum self-magnetization  satisfying
\begin{equation}
\eta= \frac{1}{1+e^{-\frac{\eta}{2 \alpha} }}.
\end{equation}
Thus, the self-magnetization avoids  the divergence of both
$\Omega_{0w}$ and ${\cal M}_{0w}$.

\begin{figure}[t]
\includegraphics[width=15.0cm]{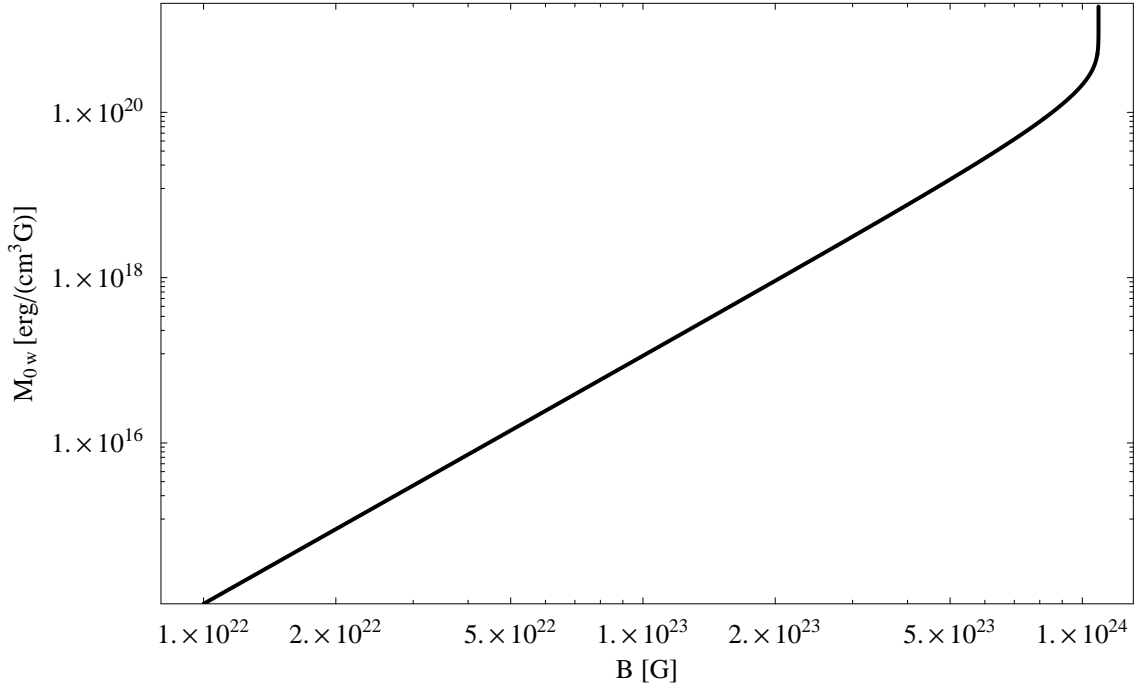}
\caption{$W$ vacuum magnetization  vs. magnetic field in a
logarithmic scale. Note that ${\cal M}_{0w}$ diverges for $B
\rightarrow B_{wc}$. A similar behavior is expected for the neutral
boson case.}
\end{figure}

Using the previous expressions (\ref{parpr}), (\ref{magnpr})  we
find a positive pressure in the direction parallel to the field
$P_{0w3}=-\Omega_{0w}$, and a negative perpendicular pressure
\begin{equation}
P_{0w \perp}=\Omega_{0w} -\frac{eB
m_{W}^{2}}{16\pi^2}\int_0^{\infty}e^{-B_wc x/B}(\frac{1+2cosh
2x}{sinh x} -\frac{3}{x}-\frac{7x}{2})\frac{ d x}{x}<0.
\end{equation}
This leads to magnetostrictive effects for  any  value of the
magnetic field since $W^{+}-W^{-}$ vacuum is compressed
perpendicularly to $B$, due to the  negative  pressures,  and  as
the pressure  $P_{0w3}$  is  positive,  it is stretched in along
$B$.

\section{Neutral vector bosons}
 It is believed that neutron stars magnetic fields could be
produced due to ferromagnetic spin coupling of neutrons. The boson
state resulting from such pairing is more favorable energetically,
since its Gibbs free energy is smaller than that of the original
neutron system \cite{herman}.
 For  neutral vector bosons with an anomalous magnetic moment
we use (from  Ref.\cite{herman}) the following spectrum
\begin{equation}\nonumber
E_{nb}(\eta)=\sqrt{p_{3}^{2}+p_{\perp}^{2}+m_{nb}^{2}+\eta
qB\sqrt{p_{\perp}^{2}+m_{nb}^{2}}} \label{neuboson}.
\end{equation}
$\eta= -1,0,1$, leading to states with magnetic moment
\begin{equation}
  \mu(\eta)=-\frac{\partial E_{nb}}{\partial
  B}\mid _{\textbf{p}=0}=\frac{-\eta q}{2\sqrt{m_{nb}+\eta
qB}}.
\end{equation}
The ground state contain again effectively massless particles,
\begin{equation}\nonumber
E_{nb}(\eta =-1,\textbf{p}=0)=\sqrt{m_{nb}^{2}- qBm_{nb}}
\end{equation}
 vanishes for
$B=B_{nbc}=\frac{m_{nb}}{q}$, and becomes imaginary for $B>B_{nbc}$,
 in analogy to the charged case, leading to the instability.

The statistical part of the thermodynamic  potential is
\begin{equation}
\Omega_{snb} = -\frac{1}{4 \pi^2 \beta} \sum_{\eta}
\int_{0}^{\infty}p_{\perp} dp_{\perp}\int_{-\infty}^{\infty} dp_3
\ln [ (1 - e^{-(E_{nb} - \mu_{nb})\beta})(1 - e^{-(E_{nb} +
\mu_{nb})\beta}) ].
\end{equation}

  The zero point energy density is
 \begin{equation}\nonumber
\Omega_{0nb} = -\frac{1}{4\pi^2}
 \int_{-\infty}^{\infty}dp_{3}
 \int_{0}^{\infty}p_{\bot}dp_{\bot}\Sigma_{\eta}E_{nb}.
\end{equation}
 By summing and integrating over all degrees of freedom and
after regularization it leads to the Euler-Heisenberg like
expression
\begin{equation}\label{NBpot}
 \Omega_{0nb} =
-\frac{(qm_{nb}B)^2}{8\pi^2}(I_{0}^{(2)}+I_{1}^{(3)}+I_{2}^{(2)})
\end{equation}
where $-\frac{(qm_{nb}B)^2}{8\pi^2}I_{0}^{2}$ is the contribution of
the states $E_{nb}(p_{\perp}=0)$, and
\begin{equation}\nonumber
  I_{0}^{(k)}=\int_0^{\infty}e^{-\frac{B'_c}{B}x}(\cosh
x-1)\frac{dx}{x^{k}},
\end{equation}
\begin{equation}\nonumber
  I_{1}^{(k)}=\int_0^{\infty}e^{-\frac{B'_c}{B}x}(\cosh
x-1-\frac{x^{2}}{2})\frac{dx}{x^{k}},
\end{equation}
\begin{equation}
 I_{2}^{(k)}=\int_0^{\infty}
\int_0^{\infty} e^{-\frac{B'_c}{B}x(u+1)^{2}}(\sinh
x(u+1)-x(u+1)-\frac{x^{3}(u+1)^{3}}{6})\frac{dud x}{x^{k}}.
\end{equation}

 The neutral boson vacuum
magnetization is
\begin{equation}\label{magnetization}
{\cal M}_{0nb}=-2\frac{\Omega_{0nb}}{B}+ \frac{q
m_{nb}^{3}}{8\pi^2}(I_{0}^{(1)}+I_{1}^{(2)}+I_{2}^{(1)})>0
  \end{equation}
The magnetization (\ref{magnetization}) is a positive quantity and
diverges for $B \rightarrow B_{nbc}=\frac{m_{nb}}{q}$, due to the
behavior of the states $E_{nb}(p_{\perp}=0)$. But this means that
neutral boson vacuum also can self-consistently maintain the field,
keeping $B<B_{nbc}$.

 Again,
vacuum paramagnetic properties conduce to the achievements of
anisotropic pressures $P_{0nb 3}=-\Omega_{0nb}>0$ and
\begin{equation}
P_{0nb \bot}=\Omega_{0nb} -\frac{qB
m_{nb}^{3}}{8\pi^2}(I_{0}^{(1)}+I_{1}^{(2)}+I_{2}^{(1)})<0.
\end{equation}

For a gas of density $N_{nb}$  in the condensate ${\cal M}_{nb}= q
m_{nb} N_{nb}/(2\sqrt{m_{nb}^2-m_{nb}qB})$, and according to the
value of $N_{nb}$, self consistent fields $B= 4\pi {\cal M}_{nb}$
may occur up to $B\sim 10^{17}$G. This might be another mechanism
for production of extremely large fields in neutron stars
\cite{herman}. For $B \sim m_{nb}/q \sim 10^{20}$G, ${\cal M}_{nb}$
diverges.




\subsection{QED vacuum}
For an electron (positron) in an external magnetic field
\begin{equation}
 E_{n}=\sqrt{ p_{3}^{2}+m_e^{2}c^4+2eB n}, \hspace{1cm} n=0,1,2...,
\end{equation}
 and the zero point energy density (Euler-Heisenberg term) in the tree
level approximation is \cite{qed1,qed2}
\begin{equation}\Omega_{0e} = \frac{e^2
B^2}{8\pi^2}\int_0^{\infty}e^{-B_{ec}
x/B}(\frac{coth x}{x} -\frac{1}{x^2}-\frac{1}{3})\frac{d x}{x}<0
\end{equation}
$B_{ec}=\frac{m_e^{2}}{e }=4.41\cdot10^{13}G$ is the critical
Schwinger field.
 \begin{equation}{\cal
M}_{0e}=-2\frac{\Omega_{0e}}{B}- \frac{e^{2}
B_{ec}}{8\pi^2}\int_0^{\infty}e^{-B_{ec} x/B}F(x)_{HE} d x>0.
\end{equation} The
electron-positron vacuum shows a paramagnetic behavior, but ${\cal
M}_{0e}\ll B$.

 But for $B\sim B_{ec}=m_{e}^2/e=4.41\times10^{13} G$, the
QED
 vacuum polarization effects, like the creation of electron-positron
 pairs
 by a photon,  become important. Photons coexist
  with mutually independent virtual $e^{+}-e^{-}$
 pairs and with bound $e^{+}-e^{-}$ virtual states (positronium),
 which is related to the singular behavior of the
 polarization operator $\Pi_{\mu\nu}$ near the thresholds for these processes\cite{AShabad}. Such singularity contributes with an absorptive term:
 vacuum becomes unstable and decays in observable $e^{+}-e^{-}$
 pairs or positronium. The first threshold for free pair creation occurs for the minimal photon energy
 $\omega= 2m_e$. For a smaller energy, $\omega =m_{p} = \lambda \cdot 2m_e$
 vacuum decay in $e^{+}-e^{-}$ bound states, i.e., positronium, where $\lambda=1- \triangle \varepsilon/2m_{e}$ and $\triangle \varepsilon$ is the positronium binding energy.
 We want to address the problem of the positronium
    vacuum in a magnetic field.

The positronium mass in the Landau ground state of the electron
and positron is $m_{p}=2m_{e}- \triangle \varepsilon$,
 where
$\triangle\varepsilon$ is the binding energy which is due to the
Coulomb interaction between the electron and the positron in
presence of a high magnetic field ($m_{p}(B) \lesssim 2 m_{e}$). For
the Coulomb ground state, $\triangle\varepsilon$ reaches high values
when the distance between the Larmor orbits of the electron and the
positron tends to zero\cite{Shabad}. We introduce as fundamental
assumption that the relativistic expression for the energy of such
one-dimensional positronium state bearing an anomalous magnetic
moment is a particular case of (\ref{neuboson})
 $E_{p}=\sqrt{p_{3}^{2}+m_{p}^{2}-qBm_{p}}$, with
 $q=2\mu_{B}$, and a degeneracy factor $eB$.

 This is equivalent to assume the bosonization of the pair
resulting from the parallel and antiparallel spin coupling of
virtual electrons and positrons, leading to a neutral boson with a
magnetic moment $q=2\mu_{B}$ and confined to move parallel to the
field $B$. These virtual positronium states lead to a logarithmic
divergence of the neutral boson vacuum magnetization for $B
\rightarrow B_{pc}\lesssim 2m_{e}^{2}/e$.

The energy density is
\begin{equation}\nonumber
  \Omega_{0p}=-\frac{(eB)^2}{2\pi^2}\int_0^{\infty}e^{-\frac{B_{pc}}{B}x}(\cosh
x-1)\frac{dx}{x^{2}}
\end{equation}
One can obtain easily the magnetization
\begin{equation}\nonumber
  {\cal M}_{0p} =-\frac{\partial\Omega_{0p}}{\partial B}
\end{equation}
It is easily seen that ${\cal M}_{0p}$ diverges logarithmically.
This divergence is indicating a phase transition to a
ferromagnetic state for $B \approx B_{pc}$. By equating $B=4\pi
{\cal M}_{0p}$ and calling $\eta'=B/B'_{pc}$, one obtain a vacuum
self-magnetization satisfying
\begin{equation}
\eta'= \frac{1}{1+e^{-\frac{\eta'}{4 \alpha} }}.
\end{equation}
This suggests the arising of a ferromagnetic phase transition for
QED vacuum, which we may understand as due to a sort of
Bose-Einstein condensation of positronium in its ground state
whose  effective mass is arbitrary small.

It means that near $B_{pc}$ all the previous considerations about
neutral boson vacuum behavior for $B \rightarrow B_{pc}$ are
applicable to the positronium vacuum; in particular, vacuum
self-magnetization might be possible in
 QED.
 \section{Conclusions}
 For electroweak vacuum, the contribution from $W^{\pm}$ vector bosons
 to
the ground state energy shows an instability for $B>B_{wc}=
m_{w}^{2}/e$.  The magnetization ${\cal M}_{w}$  diverges in both
the dense medium and the vacuum cases  for $B \to B_{wc}$. By
equating $B=4\pi{\cal M}_w$, the field can be self-consistently
maintained, i.e. becomes a ferromagnet. This mechanism actually
prevents $B$ from reaching $B_{wc}$.
 For neutral vector bosons with an anomalous magnetic moment
$q$ the ground state also shows an instability  for $B> B_{nbc}=
m_{nb}^{2}/q$ in a medium and in vacuum. ${\cal M}_{0nb}$ also
diverges for $B\rightarrow B_{nbc}$ and, as a consequence, can be
self-consistently maintained, keeping $B<B_{nbc}$. We conjecture
this mechanism might be applied to magnetized QED vacuum by assuming
virtual positronium as the neutral vector particle with anomalous
magnetic moment. Such phase transition would mean the arising of an
``order parameter", or symmetry breaking of vacuum. It is due to the
condensate of quasi-massless particles, bearing some analogy with
the Goldstone case.

\end{document}